%% file: MSSarxiv.tex
\documentclass[12pt,reqno]{article}
\usepackage{amsmath}
\usepackage{amsfonts}
\usepackage{amssymb}
\newcommand{\ra}{\rightarrow}

\def\QR{\mathbb{R}} 
\def\QC{\mathbb{C}}
\def\QZ{\mathbb{Z}} 
\def\a{\alpha}
\def\b{\beta}

\def\HRT{H_{\rm rt }}
\def\HBT{H_{\rm bt }}

\newcommand{\beq}{\begin{equation}}
\newcommand{\eeq}{\end{equation}}

\def\beas{\begin{eqnarray*}}
\def\eeas{\end{eqnarray*}}
\def\bea{\begin{eqnarray}}
\def\eea{\end{eqnarray}}
\def\a{\alpha} 
\def\b{\beta} 
 
\newcommand{\remlst}{\begin{list}
{(\arabic{num})}{\usecounter{num}\topsep0cm \itemsep0cm \parsep0cm}}
\title{\bf On Minisuperspace Models of S-branes}  
\author{\\[5mm] Stefan Fredenhagen$^1$ and Volker Schomerus$^2$ 
  \\[5mm]$^1$ CPHT -- Ecole Polytechnique,\\
F-91128 Palaiseau CEDEX, France\\[3mm] 
$^2$ Service de Physique Th{\'e}orique, CEA Saclay,\\ 
F-91191 Gif-sur-Yvette CEDEX, France\\[5mm] }
\date{August 29, 2003} 
\begin{document}
\begin{titlepage}      \maketitle       \thispagestyle{empty}

\vskip1cm
\begin{abstract} 
In this note we reconsider the minisuperspace toy models for 
rolling and bouncing tachyons. We show that the theories 
require to choose boundary conditions at infinity since 
particles in an exponentially unbounded potential fall to 
infinity in finite world-sheet time. Using standard techniques 
from operator theory, we determine the possible boundary 
conditions and we compute the corresponding energy spectra 
and minisuperspace 3-point functions. 
Based on this analysis we argue in particular that
world-sheet models of S-branes possess a discrete
spectrum of conformal weights containing both positive
and negative values. Finally, some suggestions are 
made for possible relations with previous studies of the 
minisuperspace theory.
\end{abstract} 

\vspace*{-18.5cm}\noindent
{\tt {SPhT-T03/121}} \hfill
{\tt {hep-th/0308205}}\\
{\tt {CPHT-RR-051-0803}}
\bigskip\vfill
\noindent
\phantom{wwwx}{\small e-mail: }\parbox[t]{8cm}{\small\tt
stefan@cpht.polytechnique.fr\\
vschomer@spht.saclay.cea.fr} 

\end{titlepage} 

\baselineskip=19pt 
\setcounter{equation}{0} 
\section{Introduction}
\def\tr{{\rm tr}}
Time dependent open and closed string backgrounds have received 
a lot of attention during the last year. Some of the initial 
motivation was rooted in the search for a dS/CFT correspondence 
from which the notion of S-branes emerged \cite{Gutperle:2002ai}. 
Their world-sheet description remained obscure until A.\ Sen 
proposed \cite{Sen:2002nu,Sen:2002vv} to construct them 
by adding a $\cosh X_0$-shaped boundary interaction to the free 
time-like bosonic field theory. A very closely related theory 
with a Liouville-like $\exp X_0$ boundary potential was first 
studied by Strominger \cite{Strominger:2002pc} in a minisuperspace
approximation. It describes a half-brane, i.e.\ a brane that decays 
as time evolves.
\smallskip 

Our knowledge about these two world-sheet theories is still 
quite limited. Most of the recent work centered around their 
boundary states (see e.g.\ \cite{Sen:2002nu,Sen:2002vv,
Mukhopadhyay:2002en,Okuda:2002yd,Chen:2002fp,Rey:2003xs,
Larsen:2002wc,Lambert:2003zr}). The latter were obtained 
through analytic continuation ('Wick-rotation') from Euclidean 
theories with a $\cos X$ or $\exp i X$ boundary potential, respectively. 
Other quantities, such as the boundary 2- and 3-point functions 
or bulk-boundary 2-point functions, have not been computed (some 
proposal for the boundary 2-point function can be found in 
\cite{Gutperle:2003xf}). But in the light of the analysis in 
\cite{Schomerus:2003vv}, one outcome of a full construction appears 
to be certain: most 
quantities in these two backgrounds cannot be obtained through 
Wick-rotation simply because higher correlators in the Euclidean 
models are no longer analytic.\footnote{Though it might be possible
exploit the same trick as in \cite{Schomerus:2003vv} where a non-analytic 
$c=1$ model related to closed string tachyon decay was embedded into a 
family of analytic theories with $c\geq 1$.}  
\smallskip

Unfortunately, general tools for the direct construction of 
2-dimensional world-sheet models on non-trivial time-like or 
Lorentzian space-times have not been developed. On the other 
hand, technology for the analysis of the corresponding 
minisuperspace models is certainly available and in fact fairly 
standard. Hence, investigations of such toy particle models do 
not require to pass through the corresponding Euclidean 
background. Nevertheless they can still cast some light 
on the structure of the more complicated field theory models. We 
take this as a motivation to re-investigate the time-like 
minisuperspace toy models for rolling and bouncing tachyons. 
Our analysis uncovers that the Lorentzian models have 
features which are very distinct from their Euclidean 
counterparts. In particular, parameters emerge that do 
not appear in the classical action and the spectra of 
world-sheet Hamiltonians possess new branches of discrete 
eigenvalues. 
\medskip 

In the case of the rolling tachyon theory much of the relevant 
mathematical results were previously derived in \cite{Fulop:1995di,
Kobayashi:1996kg}, 
obviously with different applications in mind. Below we shall 
review and extend these findings. Several arguments will be 
presented showing that a consistent quantum theory requires 
to choose boundary conditions in the far future. More precisely, 
wave functions $\psi$ of the rolling tachyon model
\begin{equation} \label{RTH}   
\HRT \ = \  \partial_{x_0}^2 + \lambda e^{2 x_0} \ \ 
\mbox{ with }  \ \ \ \lambda > 0 
\end{equation} 
must behave asymptotically as 
\begin{equation}  \label{RTasym} 
 \psi(x_0) \ \stackrel{x_{0} \to \infty  }{\sim} \ \text{const.} \, \cdot 
e^{-x_0/2} \cos (\sqrt{\lambda }e^{x_0}-\frac{\pi }{4}-\pi \nu_0) 
\end{equation} 
where $\nu_0 \in (0,1]$ is some real parameter. The extension of 
$\HRT$ to wave functions with such asymptotics is necessary to 
turn $\HRT$ into a consistent, self-adjoint Hamilton operator 
in the `world-sheet' theory. After we have chosen such an 
extension, we can look for eigenfunctions and determine the 
spectrum of eigenvalues $\Delta$ of the rolling tachyon model. 
Explicit formulas for the eigenfunctions are provided in 
section 2 (see eqs.\ (\ref{eig<},\ref{eig>})). The associated 
spectra depend on the boundary conditions $\nu_0$ and consist 
of a continuous and a discrete branch. More precisely, we shall 
find
$$ {\text{Spec}}_{\nu_0}(\HRT) \ = \ (-\infty ,0] \, \cup \ 
   \bigcup_{n=0}^\infty \{ 4(\nu_0+n)^2\} \ \ . 
$$   
The answer resembles the spectrum of a Schr{\"o}dinger problem 
on a half-line. Hence, one may conclude that the coordinate 
$x_0$ is somehow cut off in the future. This effect can 
actually be traced back to the structure of solutions in 
the associated classical model. In fact, it is not hard to 
see that a particle in an exponentially decreasing potential 
$V(x_0) = - \lambda e^{2x_0}$ falls to $x_0 = \infty$ in 
finite `world-sheet' time. 
\smallskip 

These findings differ significantly from the minisuperspace
analysis of Strominger \cite{Strominger:2002pc} (see also 
\cite{Kluson:2003rd,Kluson:2003xn}).
Strominger's toy model arises from a continuation of
the Euclidean theory and consequently the parameter $\nu_0$
does not appear. To investigate the relation between the two
approaches, we shall compute the minisuperspace analogues of 
the 2- and 3-point functions in our framework. Integration 
over the parameter $\nu_0$ is then shown to lead to the 
corresponding quantities in Strominger's minisuperspace 
model (see \cite{Strominger:2002pc,Schomerus:2003vv}). 
\medskip 

Similar results are also derived for the bouncing tachyon 
Hamiltonian (see e.g.\ \cite{Maloney:2003ck,Kluson:2003sh} 
for previous studies of the minisuperspace model in the context 
of tachyon dynamics) 
$$ \HBT \ =\ \partial _{x_{0}}^{2}+2\lambda \cosh 2x_{0} \ \ . $$ 
In this case, there exists a 4-parameter family of extensions,
but only two parameters remain if we require that the far 
future is not coupled to the far past. The corresponding 
wave functions possess the same asymptotics as for the 
rolling tachyon, i.e.\ 
$$ \psi (x_0)\ \stackrel{x_0 \ra \pm \infty}{\sim} 
     \ \text{const.} \ \cdot e^{-|x_0|/2}
  \cos (\sqrt{\lambda}e^{|x_0|}-\frac{\pi }{4}-\pi \nu_\pm) $$ 
with two real parameters $\nu_\pm \in (0,1]$. For all choices of 
the boundary parameters, the spectrum of eigenvalues $\Delta$ is 
purely discrete. We shall not construct it explicitly, but we shall 
be able to make two statements about its behaviour for $|\Delta| 
\rightarrow \infty$. The first one concerns the allowed eigenvalues 
for $\Delta \gg 2 \lambda$,  
$$  {\text{Spec}}_{\nu_\pm}(\HBT) \cap \{\Delta >4N^{2} \}\ \sim \ \bigcup_{n=N}^\infty 
  \{ 4(\nu_+ +n)^2\} \cup \bigcup_{n=N}^\infty \{ 4(\nu_- +n)^2\} 
  \ \ \ \mbox{ for } 
\ \ \ N \rightarrow \infty \ \ . $$ 
On the other side, when $-\Delta$ becomes very large we can show 
that the distance $\delta \Delta$ between two consecutive spectral 
lines behaves as 
$$    \delta \Delta \ \sim \ \frac{2\pi \sqrt{-\Delta }}{\log 
       (-4\Delta /\lambda )} \ \ \ \mbox{ for } \ \ \ \Delta
      \rightarrow - \infty\ \ . 
$$  
The explanation of these results and their consequences are 
similar to the discussion for the rolling tachyon background. 
\medskip 

Our presentation begins with the rolling tachyon model. 
After a brief study of its classical solutions we recall some 
simple facts from operator theory that are needed for the analysis
of the Hamiltonian in the quantum theory. We then spell out 
and prove very explicit formulas for the wave functions, the 
spectrum and the minisuperspace analogue of the 3-point 
functions. The bouncing tachyon model is addressed in section 
3. In this case, the expressions we arrive at are a little 
less explicit, but they are sufficient to establish the 
picture we have sketched above. 
 
\section{The rolling tachyon model} 

In this section we investigate the minisuperspace toy model of
a rolling tachyon background. In order to understand the
necessity of imposing boundary conditions at infinity, we 
begin with some remarks on classical solutions. Then we analyse 
the admissible boundary conditions in the quantum theory and we
determine the spectra of the associated models. Finally, we 
compute the minisuperspace analogues of the 3-point function 
and conclude with some remarks about their relation with 
quantities calculated in \cite{Strominger:2002pc,Schomerus:2003vv}. 
  
\subsection{Classical solutions of the rolling tachyon model} 

The classical action of the open string rolling tachyon background
on a Lorentzian world-sheet $\Sigma = \QR \times [0,\pi]$ takes 
the form 
$$ S^{o}_{\rm RT} \ = \ - \frac{1}{4 \pi} \int_{\Sigma}  
      d\sigma dt  
     \left(  \partial_t X_0 \partial_t X_0  -  
          \partial_\sigma X_0 \partial_\sigma X_0 \right) 
    -  {\lambda^o} \int dt \ e^{X_0}\ \ . 
$$ 
Throughout this note we set $\a'= 1$ and the constant $\lambda^o$ 
is assumed to be positive. The associated minisuperspace toy model 
is obtained by replacing the world-sheet field $X_0$ with a map 
$\tilde x_0 = \tilde x_0(t)$ which is independent of the world-sheet 
coordinate $\sigma$, 
$$ S^o_{\rm rt} \ = \ - \int_{-\infty}^{\infty}  dt \ \left(
    \frac{1}{4} \partial_t \tilde x_0 \partial_t \tilde x_0 
    +  {\lambda^o} \, e^{\tilde x_0}\right)\ \ . 
$$ 
It is not difficult to find classical solutions of this model.
They are parametrised by two real parameters $\Delta^o$ and $t_f$.
The explicit form of the solutions depends on whether $\Delta^o$ 
is positive or negative. For $\Delta^o \leq 0$ we find 
$$ e^{\tilde x_0(t)} \ = \ - \frac{\Delta^o}{\lambda^o} 
   \frac{4 \exp 2 \sqrt{-\Delta^o} (t-t_f)}
         {\left(1-\exp 2 \sqrt{-\Delta^o} (t-t_f)\right)^2} \ \ , 
$$ 
while solutions with positive $\Delta^o > 0$ can be written in 
the form 
$$ e^{\tilde x_0(t)} \ = \ \frac{\Delta^o}{\lambda^o} \sin^{-2} 
         \sqrt{\Delta^o} (t-t_f) \ \ . 
$$    
A short computation shows that both expressions correspond to 
stationary points of $S^o_{\rm rt}$. The parameters $\Delta_o$ 
and $t_f$ are fixed by the initial conditions at $t = 0$. 
$\Delta^o$ may be interpreted as the `world-sheet' energy. The 
parameter $t_f$, on the other hand, is the `world-sheet' time 
at which we reach $\tilde x_0 = \infty$. Note that for all finite 
choices of initial conditions, the time $t_f$ is finite. In other 
words, the dynamics of the rolling tachyon model implies that the 
far future $\tilde x_0 = \infty$ is reached in finite `world-sheet' 
time. Experience with similar models suggests that we shall have to 
impose some boundary condition at $\tilde x_0 = \infty$ in order 
to make the associated quantum theory well defined. We shall show 
shortly that this is indeed the case.  
\smallskip 

Let us also briefly comment on the minisuperspace model for the 
closed string rolling tachyon background. The latter is derived 
from the corresponding action 
$$ S^{c}_{\rm RT} \ = \ - \frac{1}{4 \pi} \int_{\Sigma}  
      d\sigma dt  
     \left(  \partial_t X_0 \partial_t X_0  -  
          \partial_\sigma X_0 \partial_\sigma X_0  
      +  4 \lambda \, e^{2X^0} \right) \ \ . 
$$ 
Reduction to maps $x_o(t)$ which are independent of $\sigma$ gives 
the following minisuperspace toy model for the closed string theory  
$$ S^c_{\rm rt} \ = \ -  \int_{-\infty}^{\infty}  dt \ 
    \left(\frac{1}{4} \partial_t x_0 \partial_t x_0 
    +  \lambda \, e^{2x_0}\right)\ \ . 
$$     
Note that this is essentially the same model as in the case of
open strings. In fact, the two theories can be rewritten into 
each other by means of the replacement rules $\tilde x_0 
\leftrightarrow 2 x_0$, $\lambda^o \leftrightarrow 4 \lambda$
and $\Delta^o \leftrightarrow 4 \Delta$ where $\Delta$ is the 
world-sheet energy in the closed string model. Since some of 
the formulas below are a bit simpler when written in terms of 
the closed string parameters, we shall use them from now on. 
Rewriting our results for the open string model is trivial. 

\subsection{Spectral analysis for the rolling tachyon} 

Our aim now is to study the following toy model Hamiltonian for 
the rolling tachyon background,   
$$ \HRT \ = \ \partial_{x_0}^2  + \lambda e^{2x_0} \ \ \mbox{ with } 
 \ \ \ \lambda > 0 \ \ . $$
This Hamiltonian should be regarded as the minisuperspace analogue
of the operator $L_0 + \bar L_0$ (or $L_0$ when we deal with open 
strings) in the conformal field theory of rolling tachyons. In both 
the conformal field theory and its toy model, the Hamiltonian is the
generator of `world-sheet' time translations and therefore it has to 
be be self-adjoint. 
\smallskip 
 
Our Hamiltonian $\HRT$ is originally defined on the space of smooth 
functions with compact support and while it is not self-adjoint on 
this domain, it is easily seen to admit a 1-parameter family of 
self-adjoint extensions. For each of these extensions, the spectrum 
of eigenvalues and the associated eigenfunctions have been determined 
in \cite{Fulop:1995di,Kobayashi:1996kg}. The possible eigenvalues are 
toy model analogues of the 
scaling dimensions in the conformal field theory. Their spectrum turns 
out to be continuous for negative eigenvalues.\footnote{Note that we 
have changed the sign of the Hamiltonian in comparison to 
\cite{Fulop:1995di,Kobayashi:1996kg}.} 
On the other hand, discrete positive eigenvalues can appear. 
Furthermore, there is only a single eigenfunction for each allowed 
eigenvalue, just as in spectral problems on the half-line. In this 
sense, the theory appears as if it was cut off in the far future,  
thereby reflecting the properties of the classical system that we 
discussed in the previous subsection. 
\smallskip

Even though all these results can be found in the cited literature, 
we shall briefly go through their derivation, mainly 
to prepare for the analysis of the bouncing tachyon model. Let us 
therefore begin by showing that there is indeed a 1-parameter family 
of self-adjoint extensions. According to standard operator theory 
(see e.g.\ \cite{Reed:book2}), we are supposed to determine the 
so-called deficiency indices of $\HRT$, i.e.\ we have to determine 
the number of independent square-integrable solutions for each of 
the two differential equations $\HRT \psi^\pm = \pm i \psi^\pm$. 
The problem is fairly easy to analyse. In fact, 
for each sign, the equation certainly has two linearly independent 
solutions which we can choose as  
$$ \psi^\pm_1 (x_{0}) \ = \ J_{\eta^\pm}(\sqrt \lambda e^{x_0}) 
   \ \ \ \mbox{and } \ \ \ 
   \psi^\pm_2 (x_{0}) \ = \ J_{-\eta^\pm}(\sqrt \lambda e^{x_0})
   \ \  
$$ 
with $\eta^+ = \exp(3\pi i/4)$ and $\eta^- = \exp(\pi i/4)$. Here,
$J_{\nu }$ is a Bessel function of the first kind. Among these four 
functions, two are not square-integrable because they diverge 
exponentially in the far past $x_0 \rightarrow -\infty$. Hence, 
only the functions $\psi^-_1$ and $\psi^+_2$ satisfy our search 
criteria. Since there is one such function for each sign in the 
differential equation, the deficiency indices are $(1,1)$. This 
tells us that there exists indeed a 1-parameter family of 
self-adjoint extensions as we have claimed above.           
\smallskip

We shall denote the corresponding (real) parameter by $\nu_0$ 
and the domains of the associated self-adjoint extensions by 
$D_{\nu_0}(\HRT)$. The direct approach to finding the spectrum 
of these extensions is to construct the domains $D_{\nu_{0}} 
(\HRT)$ and to diagonalize $\HRT$ thereon. Since the construction 
of the domains is a bit formal, we shall postpone this issue and 
start with a somewhat simpler and more intuitive analysis of the 
spectrum by looking for a 'maximal set of compatible 
eigenfunctions'. 

To this end, let us assume that $\HRT$ has been extended to one 
of the domains $D_{\nu _{0}} (\HRT)$. On each domain, the operator 
$\HRT$ is symmetric, i.e.\ 
$$ \langle \psi| \HRT \psi'\rangle \ = \ 
   \langle \HRT \psi| \psi'\rangle 
   \ \ \ \mbox{ for all } \ \ \ \ 
    \psi,\psi' \in D_{\nu_0}(\HRT) \ \ .
$$ 
If we insert the explicit form of $\HRT$, this property can be 
re-expressed as a boundary condition for functions in the 
domain, 
\begin{equation} \label{symprop} 
 \big(\  \overline \psi (x_0) \, \partial_{x_0} \psi'(x_0) 
   - \psi'(x_0) \, \partial_{x_0} \overline \psi(x_0)\,  
    \big) \big|_{-\infty}^\infty 
   \ = \ 0 \ \  \  \mbox{ for all } \ \ \ \ 
    \psi,\psi' \in D_{\nu_0}(\HRT) \ \ . 
\end{equation} 
The different domains we can select for self-adjoint extensions
of $\HRT$ correspond to different choices of boundary conditions 
at infinity. All of these boundary conditions have to be consistent 
with the previous equation. This provides us with some sort of 
compatibility condition which two functions from the same domain
$D_{\nu _{0}} (\HRT)$ have to satisfy.     
\smallskip 

We can use this condition to find sets of compatible eigenfunctions 
$\psi^{\nu_{0}}_{\Delta }$ of $\HRT$, i.e.\ compatible solutions of 
the differential equation  
\begin{equation} \label{Liouveq}   
\HRT \, \psi^{\nu_0}_\Delta (x_0) \ = \  \Delta \, \psi^{\nu_0}_\Delta(x_0) 
\ \ \ \mbox{ for }\  \ \ \ \psi^{\nu_0}_\Delta \in D_{\nu_0}(\HRT)\ \ \ .
\end{equation} 
As we shall see momentarily, the boundary condition (\ref{symprop}) 
applied to an arbitrary pair of eigenfunctions turns into a powerful 
constraint, both on the allowed eigenvalues and on the form of the 
eigenfunctions.
\smallskip  

To begin with, let us study the situation where $\Delta = 4 \nu^2$ 
is positive. In this case, one of the Bessel functions that solves
the differential equation (\ref{Liouveq}) diverges exponentially 
in the far past so that eigenfunctions of $\HRT$ are necessarily 
of the form 
\begin{equation} \label{eig<} 
 \psi^{\nu_0}_\nu (x_0) \ = \ 2 \sqrt{\nu }\ J_{2\nu} 
(\sqrt{\lambda} e^{x_0}) 
\ \ \mbox{ for } \ \nu > 0   
\end{equation} 
where the prefactor is chosen s.t.\ $\|\psi^{\nu _{0}}_{\nu }\|_2=1$.
Using the asymptotic behaviour of Bessel functions for large 
arguments, it is then easy to see that the condition (\ref{symprop})
for two such functions $\psi _{\nu }^{\nu_0},\psi _{\nu '}^{\nu
_{0}}$ to be in the domain of the same self-adjoint extension becomes 
$$ \sin \pi (\nu - \nu') \ = \ 0 \ \ .$$
This means that $\nu - \nu' \in \QZ$ or, equivalently, that 
the only allowed values for $\nu$ are given by $\nu_n = n + \nu_0, 
n = 0,1,2,\dots$. Here we have associated our parameter $\nu_0 \in (0,1]$ 
with the smallest positive eigenvalue that can appear. Our argument 
shows that for positive $\Delta$, the spectrum of $\HRT$ on the domain 
$D_{\nu_0}(\HRT)$ is discrete with eigenvalues of the form $\Delta_n = 
4 (n+\nu_0)^2$.
\smallskip 

Now we want to extend the analysis to include negative eigenvalues. 
In evaluating our condition (\ref{symprop}) we encounter a potential 
problem because $\psi_{\Delta} (x_{0})$ with $\Delta \leq 0$ approaches 
a free wave for $x_{0}\to -\infty $: it is not square-integrable and 
thus not in $D_{\nu _{0}} (\HRT)$. The issue is resolved by smearing 
eigenfunctions with a smooth energy distribution $\rho(\Delta)$.
No matter how sharply peaked this distribution is, the resulting 
smeared wave function and its derivative will vanish for $x_{0}\to -\infty$ 
due to the fact that the eigenfunctions oscillate faster and faster 
in the parameter $\Delta$ as we approach the far past $x_0 \rightarrow 
- \infty$. On the other side, for $x_{0}\to \infty $ the eigenfunctions 
do not show this oscillating behaviour in $\Delta $ so that their 
asymptotics can be approximated arbitrarily closely by smeared wave 
functions. This means that we are allowed to evaluate the condition 
(\ref{symprop}) at $x_{0}=\infty $ also for eigenfunctions   
associated to the negative spectrum.
\smallskip

\begin{figure}[t]
\begin{center}
\input{rt.pstex_t}
\end{center}
\caption{\label{fig:rt}An illustration of the spectrum of the rolling
tachyon model with $\lambda =0.2$ and boundary parameter
$\nu _{0}=3/4$. For $\Delta >0$ the spectrum is discrete and we
plotted the wave functions $\psi ^{\nu _{0}}_{\nu _{n}}$ for $n=0,1$.
The spectrum is continous for $\Delta <0$ and two wave functions $\psi
^{\nu _{0}}_{\omega }$ are shown as representatives.}
\end{figure}
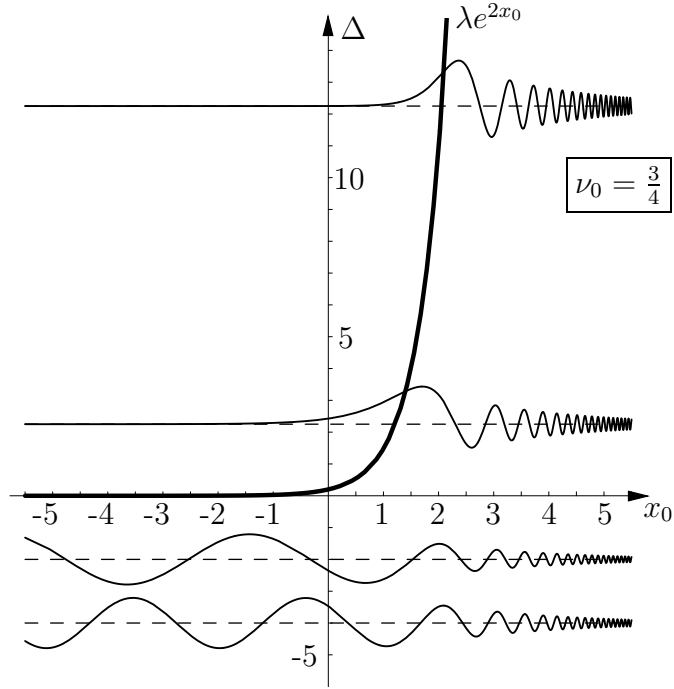

Let us now choose $\psi = \psi^{\nu_0}_{\nu_n}$ and let
$\psi'_{\Delta}$ be an eigenfunction for non-positive eigenvalue
$\Delta$. For the latter, we make the general Ansatz
$$ \psi'_\Delta(x_0) \ = \ \a_\Delta J_{-i\sqrt{-\Delta}}
   (\sqrt{\lambda} e^{x_0}) + \b_\Delta J_{i\sqrt{-\Delta}}
    (\sqrt{\lambda} e^{x_0}) 
  \ \ . 
$$ 
Applying the condition (\ref{symprop}) to $\psi $ and $\psi '$ 
then gives 
$$  
  \alpha_\Delta \sinh \pi (\omega - i\nu_0) \ = \ 
  \beta_\Delta \sinh \pi (\omega +  i \nu_0) \ \  
$$ 
where $\Delta = - 4 \omega^2$. This implies that there exists only one 
eigenfunction for each value of $\Delta \leq 0$. We can take it  
to be of the form 
\begin{equation} \label{eig>} 
 \psi^{\nu_0}_\omega (x_0) \ = \ \big( 2\omega/ \sinh 2\pi
\omega \big) ^{\frac{1}{2}} 
  \ \left( J_{-2i\omega} (\sqrt{\lambda}e^{x_0}) 
   + \frac{\sinh \pi (\omega -i\nu_0)}{\sinh \pi (\omega + i\nu_0)}
     \, J_{2i\omega}(\sqrt{\lambda} e^{x_0} )\right) \ \ .  
\end{equation}  
It is finally simple to check the condition (\ref{symprop}) at
$x_{0}=\infty$ for any pair of functions (\ref{eig>}). This is in 
fact equivalent to the orthogonality of the system of eigenfunctions 
$\psi _{\omega }^{\nu_{0}}$, 
\[
\int _{-\infty }^{\infty }dx_{0}\ \overline{\psi _{\omega }^{\nu _{0}}} 
(x_{0}) \, \psi _{\omega '}^{\nu _{0}} (x_{0})\  =\ \delta (\omega -\omega ') 
\ \ .\]
We thus arrive at a family of 'maximal sets of compatible
eigenfunctions'. The spectrum and the corresponding wave functions
are sketched in figure~\ref{fig:rt}. To be sure that we did not overlook any
part of the spectrum, we can verify completeness by showing that  
\begin{equation}\label{completeness}
\int_{0}^{\infty }d\omega \ \overline{\psi ^{\nu _{0}}_{\omega }}
(x_{0})\, \psi ^{\nu _{0}}_{\omega } (x_{0}')\ +\ \sum_{n=0}^{\infty }
\overline{\psi _{\nu _{n}}^{\nu _{0}}} (x_{0})\, \psi _{\nu _{n}} ^{\nu
_{0}} (x_{0}') \ =\ \delta (x_{0}-x_{0}') \ \ .
\end{equation}
While the proof of the orthogonality is left as an exercise, 
details on the derivation of eq.\ (\ref{completeness}) can be 
found in appendix A. We would also like to stress that all the 
eigenfunctions we found for a given parameter $\nu_0$ possess 
the same asymptotics (\ref{RTasym}) in the far future. 
\smallskip

This finally brings us back to the direct construction of the domains 
$D_{\nu_{0}} (\HRT)$ involving the functions $\psi _{1}^{-}$ and 
$\psi_{2}^{+}$. The general prescription due to J.\ von Neumann
\cite{Neumann:1929} is to first take the domain $D (\bar{H}_{\text{rt}})$ 
of the closure $\bar{H}_{\text{rt}}$ of the original operator $\HRT$ 
defined on the smooth functions with compact support. Then the domain 
$D_{\nu '} (\HRT)$ of an extension labeled by some parameter $\nu' 
\in (0,1]$ is given by 
\[
D_{\nu '} (\HRT)\ =\ \{\psi +\alpha (\psi _{1}^{-}+e^{2\pi i \nu
'}\psi _{2}^{+})|
\psi \in D (\bar{H}_{\rm rt}) \ ,\ \alpha \in \QC \} \ \ .
\]
It can be seen that $D (\bar{H}_{\rm rt})$ does not contain any of 
the eigenfunctions of $\HRT$. In fact, the latter fall off like 
$e^{-x_{0}/2}$ for $x_{0}\to \infty $ while functions in the set 
$D(\bar{H}_{\rm rt})$ decay faster. The asymptotic behaviour of 
functions in $D_{\nu'} (\HRT)$ is thus governed by the chosen 
linear combination $\psi _{1}^{-}+e^{2\pi i \nu'}\psi _{2}^{+}$. 
To determine the spectrum of $\HRT$, we only have to analyse which 
solutions of eq.\ (\ref{Liouveq}) have this specific asymptotic 
behaviour parametrised by $\nu '$. The result agrees with the 
analysis we presented before and there is a one-to-one mapping 
from $\nu _{0}$ to $\nu'$.

\subsection{Correlation functions for the rolling tachyon} 
In this subsection we want to derive expressions for the minisuperspace
analogue of 2- and 3-point functions in the conformal field theory. We 
shall begin with the 3-point functions and then recover the simpler 
2-point function as a special case. To make contact with standard conventions
in Liouville field theory we choose a different normalization for 
our eigenfunctions with $\Delta \leq 0$, 
\[
\tilde{\psi} ^{\nu _{0}}_{\omega } (x_{0})\ =\ 
(\lambda/4)^{i \omega} \Gamma(1-2i\omega) \left( J_{-2i\omega}
(\sqrt{\lambda}e^{x_0})  
   + \frac{\sinh \pi (\omega -i\nu_0)}{\sinh \pi (\omega + i\nu_0)}
     \, J_{2i\omega}(\sqrt{\lambda} e^{x_0} )\right) \ \ . 
\]
These wave functions possess the following asymptotic behaviour for 
$x_{0}\to -\infty $ 
\begin{equation}
\label{Rampl} 
\tilde{\psi } ^{\nu _{0}}_{\omega }\ \sim \ e^{-2i\omega x_{0}} +
R^{\nu_0}_{0}(\omega) \  e^{2i\omega x_{0}} \ \mbox{ ; }\  
R^{\nu_0}_0(\omega) \ = \ \left(\frac{\lambda}{4}\right)^{2i\omega} 
 \frac{\Gamma(1-2i\omega)}{\Gamma(1+2i\omega)} 
 \frac{\sinh \pi (\omega - i \nu_0)}{\sinh \pi (\omega + i \nu_0)}   
\ .
\end{equation} 
For obvious reasons, the quantity $R^{\nu_0}_0(\omega)$ is usually 
referred to as the {\em reflection amplitude}.   
\smallskip 

We can now calculate the three-point function in the minisuperspace
approximation, 
\[
C^{\nu_0}_0(\omega_1,\omega_2,\omega_3) \ = \ 
\int_{-\infty }^{\infty }dx_{0}\ \tilde{\psi} _{\omega _{1}}^{\nu
_{0}} (x_{0})\  e^{-2i\omega _{2}x_{0}}\  \tilde{\psi }_{\omega _{3}}^{\nu
_{0}} (x_{0}) \ = \ (\lambda /4)^{2i\tilde{\omega }}\  P_{0}^{\nu _{0}} (\omega
_{j})\  e^{Q_{0} (\omega _{j})}
\]
where
\[
\exp Q_{0} (\omega _{1},\omega _{2},\omega _{3})\ =\ \frac{1}{\Gamma
(1+2i\tilde{\omega })}\ \prod _{j=1}^{3}\frac{\Gamma (1+
(-1)^{j}2i\omega _{j})}{\Gamma (1- (-1)^{j}2i\tilde{\omega }_{j})}
\]
and
\[
P_{0}^{\nu _{0}} (\omega _{1},\omega _{2},\omega _{3})\ =\ 
\frac{i\pi }{2}\left[\frac{1}{\sinh 2\pi \tilde{\omega }}+
\frac{\zeta^{\nu_0}(\omega _{1})}{\sinh 2\pi \tilde{\omega }_{1}}-
\frac{\zeta^{\nu_0} (\omega_{1})\zeta^{\nu_0} (\omega _{3})}
{\sinh  2\pi \tilde{\omega }_{2}}+\frac{\zeta^{\nu_0}
(\omega _{3})}{\sinh 2\pi \tilde{\omega }_{3}} \right] \ \ .
\]
Here, we denoted by $\zeta^{\nu_0} (\omega )$ the relative phase in our
combination of the two Bessel functions,
\begin{equation}\label{defofzeta}
\zeta^{\nu_0} (\omega )\ =\ \frac{\sinh \pi (\omega -i\nu _{0})}{\sinh \pi
(\omega +i\nu _{0})} \ \ 
\end{equation}
and we introduced the quantities $\tilde{\omega },\tilde{\omega }_{j}$ 
through  
\begin{align*}
2 \tilde{\omega }\  &= \ \omega _{1}+\omega _{2}+\omega
_{3} & \tilde{\omega }_{j} \ &= \ \tilde{\omega }-\omega _{j} \ \ .
\end{align*}
As explained in \cite{Schomerus:2003vv} we can recover the 2-point function 
of the minisuperspace model from $C^{\nu_0}_0$ by sending $\omega_2$ to 
zero, 
$$ \lim_{\omega_2 \rightarrow 0} C^{\nu_0}_0(\omega _{1}, \omega _{2},
\omega _{3}) \ = \ \pi\  R^{\nu_0}_0(\omega_1) \, \delta (\omega_1 - \omega_3)
$$ 
where $R^{\nu_0}_{0}(\omega)$ was defined in eq.\ (\ref{Rampl}) above and 
we assume $\omega_1,\omega_3 \geq 0$ as before. A comparison with the 
corresponding formulas in \cite{Schomerus:2003vv} shows that the 
features of our 
minisuperspace analysis are quite distinct from the minisuperspace 
model proposed by Strominger. In fact, we recover quantities of the 
latter by replacing our phases $\zeta^{\nu_0}(\omega)$ with the real 
function $[-\exp (-2\pi \omega)]$. It might be interesting to observe 
that this replacement can be thought of as resulting from an integration 
over the parameter $\nu_0$. More precisely, using the integral 
formula 
\[
\int_{0}^{1}d\nu_{0} \ \prod _{j=1}^{n}\ \zeta ^{\nu _{0}} (\omega _{j})
\ = \ \prod _{j=1}^{n}\ \big(-e^{-2\pi \omega _{j}} \big) \quad
\text{for } \omega _{j}>0 \ \ ,
\]
one can show that $C_0 = \int d\nu_0 C^{\nu_0}_0$ and $R_0 = 
\int d \nu_0 R^{\nu_0}_0$ agree with the 3- and 2-point functions
of Strominger's minisuperspace model for the rolling tachyon (see 
\cite{Schomerus:2003vv}). The latter was obtained from the minisuperspace 
theory of the usual Liouville model by a Wick rotation.  

\section{The bouncing tachyon model} 

We now proceed to the minisuperspace analysis of the bouncing tachyon 
model. Our discussion starts with a few remarks on the classical 
theory before we enter the spectral analysis of the bouncing tachyon 
Hamiltonian. Though our formulas will not be as explicit as in the 
case of the rolling tachyon model, we shall be able to argue that 
there is a 2-parameter family of self-adjoint extensions which are 
adequate for the physics we want to study. All these extensions
possess a discrete spectrum, both for positive and for negative 
eigenvalues. 

\subsection{On the classical physics of bouncing tachyons}

The minisuperspace toy model for the bouncing tachyon background
is obtained from the corresponding 2-dimensional field theory as 
in the case of the rolling tachyon. In terms of closed string 
parameters, the classical action reads 
$$ S^c_{\rm bt} \ = \ - \int_{-\infty}^{\infty}  dt \ 
    \left(\frac{1}{4}\partial_t x_0 \partial_t x_0 
    + 2 \lambda \, \cosh 2x_0 \right)\ \ . 
$$ 
It is identical to the minisuperspace toy model for strings in an 
S-brane background up to a trivial change of parameters (see our 
discussion in section 2.1). 
\smallskip 

The key features of the classical theory can be deduced 
immediately from our discussion of the rolling tachyon model. 
Most importantly, it is easy to see that any solution of the 
bouncing tachyon theory reaches either $x_0 = -\infty$ or $x_0 
=  \infty$ in finite `world-sheet' time $t_f$. This property of 
the bouncing tachyon theory suggests that the quantum mechanical 
model requires to fix boundary conditions in the far past {\em and} 
the far future. We shall prove this in the next subsection. Moreover, 
because the model now appears to be cut off at both sides, we expect 
the spectrum to be discrete even for $\Delta \leq 2\lambda$. Since 
this is a new feature of the bouncing tachyon model, we would like 
to explore it more qualitatively and derive an explicit expression 
for the semi-classical spectral density. 
\smallskip 
     
The basic observation is that the space of classical orbits with 
energies between $\Delta _{0}$ and $\Delta $ has a finite volume 
$\Gamma (\Delta_{0},\Delta )$ . We expect the number $N (\Delta _{0},
\Delta )$ of quantum mechanical states in the energy interval 
$(\Delta_{0},\Delta)$ to be the corresponding classical volume 
in phase-space divided by $h=2\pi \hbar = 2\pi$,\footnote{Throughout 
this paper we use units in which $\hbar = 1$.}  
\begin{equation}\label{semclassrule}
N (\Delta _{0},\Delta )\ \approx \ \frac{1}{2\pi }\Gamma (\Delta
_{0},\Delta ) \ \ .
\end{equation}
In the following we keep $\Delta _{0}$ constant as a reference energy
and investigate the dependence of the volume $\Gamma $ on $\Delta $. 
For energies $\Delta_{0} <\Delta <2\lambda $ the volume in phase-space
is given by  
\[
\Gamma_- (\Delta _{0},\Delta )\ =\ 2\int_{-\infty }^{\infty }
\left(\sqrt{2\lambda
\cosh 2x_{0}- \Delta_{0}}-\sqrt{2\lambda \cosh 2x_{0}- \Delta} 
\right)dx_{0} \ \ .
\]
The integral is not difficult to evaluate and we obtain 
\begin{equation}\label{psvolnegdelta}
\Gamma_- (\Delta _{0},\Delta )\ =\ c (\Delta_{0})
+4\sqrt{2\lambda -\Delta}\left(E
\left(\sqrt{\frac{\Delta + 2\lambda }{\Delta - 2\lambda }}
\right)-K\left(\sqrt{\frac{\Delta +2\lambda }{\Delta -2\lambda }} 
\right) \right)\ \ \ . 
\end{equation}
Here, $c$ is independent of $\Delta$ and therefore it is irrelevant 
for our purposes. $K$ and $E$ are the complete elliptic integrals of 
first and second order, respectively. For any pair $\Delta_{0},\Delta 
> -\infty$ this volume $\Gamma$ is finite. A similar calculation can 
be performed for $\Delta >2\lambda $. In this case we find
\begin{equation}\label{psvolposdelta}
\Gamma_+ (\Delta _{0},\Delta )\ =\ c (\Delta_{0}) 
+4\sqrt{2\lambda+\Delta  }\ E\left(\sqrt{1-\frac{\Delta -2\lambda
}{\Delta +2\lambda }} \right) \ \ . 
\end{equation}
The dependence of the phase-space volume on the energy $\Delta $ is
shown in figure~\ref{fig:psv}.
\smallskip

Semi-classical quantities are expected to provide good approximations
for the exact quantum theory when $|\Delta|$ becomes large. Hence, we 
shall use our formulas to find the density of states for $|\Delta| \gg 
\lambda  $. In the limit $\Delta \to -\infty $ we can estimate the 
dependence (\ref{psvolnegdelta}) of the volume $\Gamma_-$ on $\Delta$ 
by  
\[
\Gamma_- (\Delta _{0},\Delta )\ =\ c (\Delta _{0})
-2\sqrt{-\Delta }\big(\log (-4\Delta/\lambda ) -2  \big)
+\mathcal{O}\bigg(\frac{\log (-\Delta) }{\sqrt{-\Delta
}}\bigg) \ \ .
\]
From the semi-classical rule (\ref{semclassrule}) we may read off the
size $\delta \Delta$ of the interval in which we find one quantum 
mechanical state, i.e.\ the level spacing
\begin{equation} 
\delta \Delta \ \approx  \ 2\pi \bigg(\frac{d}{d\Delta }\Gamma_- (\Delta
_{0},\Delta )  \bigg)^{-1} 
\label{semclassspectrum}
\ \approx  \ \frac{2\pi \sqrt{-\Delta }}{\log (-4\Delta /\lambda )} \ \ 
\ \mbox{for} \ \ \Delta \ll -\lambda\ .
\end{equation}
Our quantum mechanical models will later be shown to reproduce this
result. A similar analysis carries through in the limit where $\Delta 
\to + \infty$. In fact, for large $\Delta $ we can approximate the value 
(\ref{psvolposdelta}) of the phase-space volume $\Gamma_+$ by
\[
\Gamma_+ (\Delta _{0},\Delta )\ =\ c (\Delta _{0})
+2\pi \sqrt{\Delta }
+\mathcal{O}\bigg(\frac{1}{\sqrt{\Delta}}\bigg) \ \ .
\]
The level spacing is therefore 
\begin{equation}\label{levspacposdelta}
\delta \Delta \ \approx \ 2\sqrt{\Delta } \ \ \ \mbox{ for } \ \ 
\Delta \gg \lambda \ \ . 
\end{equation}
This concludes our analysis of the semi-classical limit for the 
bouncing tachyon model.

\subsection{Spectral analysis of the bouncing tachyon} 

In the following subsection we proceed to the quantum theory of the 
bouncing tachyon toy model, i.e.\ we investigate the 1-dimensional 
Hamilton operator
\begin{equation}\label{bthamop}
H_{\text{bt}}\ =\ \partial _{x_{0}}^{2}+2\lambda \cosh 2x_{0} \ \ . 
\end{equation}
As we have argued on the basis of the classical theory, we have to 
impose certain asymptotic boundary conditions on the wave functions. 
We shall see that when defined on the smooth functions with compact 
support, the Hamilton operator is not essentially self-adjoint. 
Instead, there is a 4-parameter family of self-adjoint extensions. 
In all of these extensions we have a purely discrete spectrum. Out 
of the 4-parameter family of extensions we are only interested in
those where the boundary condition in the far future does not feed 
particles back into the far past. The precise formulation of this 
intuitive idea is shown below to single out a 2-parameter family 
of extensions, one parameter being associated to each boundary.
\smallskip

In the quantum mechanical problem the main task is to find appropriate
domains of the Hamilton operator $H_{\text{bt}}$. Let us first show that 
on smooth functions with compact support the operator is not essentially 
self-adjoint. We determine its deficiency indices by looking for 
square-integrable solutions $\psi$ of the differential equations 
$H_{\text{bt}}\psi =\pm i \psi$. For each sign there are two linearly 
independent solutions. The asymptotics of these solutions for 
$x_{0}\to \pm \infty $ are the same as for the rolling tachyon in the 
limit $x_{0}\to \infty $, i.e.\ they fall off with an exponential 
suppression $\sim e^{-|x_{0}|/2}$. Obviously, these functions are 
square-integrable and our deficiency indices are $(2,2)$. Standard 
operator theory then tells us that there is a family of self-adjoint 
extensions $D_{U}(H_{\text{bt}})$ labeled by the set of unitary 
$2\times 2$-matrices $U$. Such matrices are parametrised by four 
independent parameters.     
\smallskip

For physical reasons we are not interested in all possible extensions. 
Since we interpret $x_0 = \pm \infty $ as far past and far future, we 
want independent conditions in the two limits. This means to replace 
a boundary condition of the form (\ref{symprop}) by two separate 
conditions, one for each boundary  
\begin{equation} \label{btsymprop} 
\lim_{x_{0}\to \pm \infty }\big(\  \overline \psi (x_{0}) \, \partial_{x_{0}} 
\psi'(x_{0}) - \psi'(x_{0}) \, \partial_{x_{0}} \overline \psi(x_{0})  
 \,  \big) \ = \ 0 \ \  \  \mbox{ for all } \ \ \ \ 
    \psi,\psi' \in D_{U}(H_{\text{bt}}) \ \ . 
\end{equation} 
Obviously, these new requirements on the asymptotics of wave functions 
are stronger than the symmetry condition (\ref{symprop}). In fact, they 
reduce the number of parameters from four to two. 
\smallskip

In finding the possible extensions, we can profit from our experience
in the rolling tachyon scenario. There, we had to include functions
$\psi$ into the domain of the Hamiltonian which behave asymptotically 
as 
\begin{equation}\label{asympbeh}
\psi (x_{0})\ \sim \ \text{const.} \ \cdot e^{-x_{0}/2}\cos (\sqrt{\lambda
}e^{x_{0}}-\frac{\pi }{4}-s) 
\end{equation}
with some shift $s$. In the rolling tachyon model this parameter had 
to be the same real number $s=\pi \nu _{0} $ for all functions in the 
domain. Here we obtain a similar result: if we want to include functions 
with an asymptotic behaviour given in (\ref{asympbeh}) for $x_{0}\to 
\infty$, it follows from eq.\ (\ref{btsymprop})that the shift $s=\pi
\nu _{+}$ has to be real and the same for all functions. This applies
analogously for the asymptotics at $x_{0}\to -\infty $ with a possibly
different boundary parameter $\nu_{-}$. We thus find domains $D_{\nu
_{+},\nu _{-}} (H_{\text{bt}})$ on which $H_{\text{bt}}$ is symmetric.
\smallskip

It is now fairly easy to see that $H_{\text{bt}}$ on such a domain is
essentially self-adjoint. What we have to decide is whether there are 
functions solving the differential equations $H_{\text{bt}}\psi^\pm =
\pm i \psi^\pm $ which are in the domain of the adjoint operator, i.e.\ 
whether there exist solutions $\psi^\pm$ satisfying 
$$ (H_{\text{bt}}\psi^\pm ,\phi ) \ = \ (\psi^\pm ,H_{\text{bt}} 
  \phi )  
\ \ \ \mbox{ for all } \ \ \  \phi \in D_{\nu _{+},\nu _{-}} 
(H_{\text{bt}})\ \ . $$
From this condition we find that such $\psi^\pm$ have to have the 
same asymptotic behaviour as the functions in the domain $D_{\nu_{+},
\nu_{-}} (H_{\text{bt}})$. It is not hard to see that these 
asymptotics are incompatible with the differential equations 
for $\psi^\pm$. Hence, the deficiency indices of the bouncing 
tachyon Hamiltonian on our extended domains are zero.
\smallskip

Having found the right domains, we can start to analyse the spectrum. 
The differential equation we have to solve when we look for
eigenfunctions $\psi_{\Delta} (x_{0})$ of (\ref{bthamop}) 
is known under the name 'modified Mathieu equation' (see e.g.\
\cite{Meixner:book}), 
\[
\psi_{\Delta} '' (x_{0}) +2\lambda \cosh (2x_{0}) \psi_{\Delta }
(x_{0})\ =\ \Delta\, \psi_{\Delta } (x_{0}) \ \ .
\]
Its solutions are given by the modified Mathieu functions of the 
first kind $\text{Ce}_{\nu} (x_{0},\lambda )$ and $\text{Se}_{\nu } 
(x_{0},\lambda )$ which are even and odd under $x_{0}\to -x_{0}$, 
respectively. The parameter $\nu =\nu (\Delta,\lambda )$ is 
called a characteristic exponent to $\Delta$ and $\lambda $. 
Note that the statements we have just made are only true if 
$\nu $ is not an integer. Since the whole story gets a lot more 
involved for integer characteristic exponent $\nu$ (see e.g.\
\cite{Meixner:book}), we assume throughout our discussion that 
$\nu $ is non-integer. For large values of $|\Delta|$ the 
characteristic exponent $\nu$ is related with the eigenvalue 
$\Delta$ by  $\Delta \approx \nu^{2}$. To  analyse the asymptotic 
behaviour of the eigenfunctions, the  following linear combinations 
are convenient 
\[
    \text{Me}_{\pm \nu } (x_{0},\lambda )\ =\ 
    \text{Ce}_{\nu } (x_{0},\lambda )\pm
    \text{Se}_{\nu } (x_{0},\lambda ) \ \ .
\]
When $|x_{0}|$ is large, they behave like Bessel functions,
\begin{align*}
\text{Me}_{\nu } (x_{0},\lambda )\ \sim & \ \frac{\text{Me}_{\nu } 
(0,\lambda)}{\text{M}^{(1)}_{\nu } (0,\lambda )} J_{\nu } 
(\sqrt{\lambda}e^{x_{0}})\quad \text{for }\ \  x_{0}\to \infty \\
\text{Me}_{\nu } (x_{0},\lambda )\ \sim & \ \frac{\text{Me}_{\nu } 
(0,\lambda)}{\text{M}^{(1)}_{-\nu } (0,\lambda )} J_{-\nu } 
(\sqrt{\lambda}e^{|x_{0}|})\quad  \text{for }\ \  
x_{0} \to -\infty \ \ .
\end{align*}
Expressions for $\text{M}^{(1)}_{\pm \nu } (0,\lambda )$ can be 
found in \cite[Chapter~2]{Meixner:book}. Here we shall only need 
the approximation for the ratio
\begin{equation}\label{approxxi}
\xi_{\nu } (\lambda  )\ :=\ \frac{\text{M}^{(1)}_{\nu } (0,\lambda
)}{\text{M}^{(1)}_{-\nu } (0,\lambda )} \ =\ 
\bigg(\frac{\lambda}{4}\bigg)^{\nu }\frac{\Gamma (1-\nu )}{\Gamma
(1+\nu )} \ \bigg(1+\mathcal{O} (\lambda ^{2}/\nu)  \bigg) \ \ .
\end{equation}
Note that this approximation is not valid when $\nu $ approaches an
integer number.\footnote{If $\nu$ is real, the error estimate in eq.\
(\ref{approxxi}) is still correct if we keep the (non-zero) fractional
part of $\nu $ fixed while letting $|\nu |$ grow.} 
Having gathered some background information we are now prepared to 
study eigenfunctions of our Hamiltonian. For such eigenfunctions 
we make an Ansatz of the form
\[
\psi_{\nu } (x_{0},\lambda )\ =\ a \text{Me}_{\nu } (x_{0},\lambda
)+b \text{Me}_{-\nu } (x_{0},\lambda )\ \ . 
\]
Using such an Ansatz, our boundary conditions with labels $\nu _{+},
\nu _{-}$ turn into two constraints for the three parameters $a,b$ 
and $\nu $,
\begin{align*}
a \sin \pi \big(\nu _{+}-\frac{\nu }{2}\big) + b\, \xi_{\nu } (\lambda
) \sin \pi \big(\nu
_{+}+\frac{\nu }{2}\big) \ =&\ 0\\
a\, \xi_{\nu } (\lambda  )\sin \pi \big(\nu _{-}+\frac{\nu }{2}\big)+b\sin\pi \big(\nu
_{-}-\frac{\nu }{2}\big)\ =&\ 0 \ \ .
\end{align*}
A non-trivial solution of these constraints exists whenever the 
following equation is fulfilled, 
\begin{equation}\label{btspectrum}
\sin\pi \big(\nu _{+}-\frac{\nu }{2}\big)\sin \pi \big(\nu
_{-}-\frac{\nu }{2}\big)\
=\ \xi_{\nu } ^{2} (\lambda  )\sin \pi \big(\nu _{+}+\frac{\nu }{2}\big)\sin \pi \big(\nu
_{-}+\frac{\nu }{2}\big)\ \ .
\end{equation}
This condition determines the spectrum of $H_{\text{bt}}$. It can 
only be satisfied for isolated values of $\nu$ and therefore we 
find a discrete spectrum. The spectrum and the corresponding wave
functions are sketched in figure~\ref{fig:bt}. For generic values of
$\nu _{+},\nu _{-}$, eq.\ (\ref{btspectrum}) cannot be solved explicitly. 
Our aim now is to determine the spectrum approximately in the limit
$|\Delta| \to \infty $. 
\begin{figure}
\begin{center}
\begin{minipage}[t]{180pt}
\input{psv.pstex_t} 
\caption{\label{fig:psv}Energy $\Delta $ versus phase-space volume
$\Gamma (0,\Delta )$. Neighbouring grid lines correspond to a 
spacing $\delta \Gamma =h$, i.e.\ in the energy interval $\delta
\Delta $ between two horizontal grid lines we expect to find one
quantum mechanical state.}
\end{minipage}
\hfill 
\begin{minipage}[t]{253pt}
\input{bt.pstex_t}\caption{\label{fig:bt}An illustration of the spectrum of
the bouncing tachyon model. The boundary labels are $\nu _{+}=3/4$ and
$\nu _{-}=1/4$, and we set $\lambda =0.2$. The spectrum is purely
discrete, and the level spacing agrees well with the semi-classical
expectations as we can see by comparing with the neighbouring
figure~\ref{fig:psv}. The right half of the drawing resembles strongly
the figure of the rolling tachyon model (fig.\ \ref{fig:rt}).}
\end{minipage}
\end{center}
\end{figure}
\smallskip

Let us first consider the case $\Delta \to +\infty $, i.e.\ $\nu
\to \infty $. Based on our semi-classical analysis in section~3.1 
we expect an average level spacing of $\delta \nu =1$ (cf.\ eq.\
(\ref{levspacposdelta})). In the exact quantum theory we conclude   
from eq.\ (\ref{approxxi}) that $\xi_{\nu } (\lambda  )$ vanishes in the limit
$\nu\to \infty$, at least when $\nu$ is not close to integers. Hence,
the condition (\ref{btspectrum}) turns into the simple relation 
\[
\sin \pi \big(\nu _{+}-\frac{\nu }{2}\big) \sin \pi \big(\nu
_{-}-\frac{\nu }{2}\big) \ =\ 0
\]
which has the solutions
\begin{equation}\label{btlargeenergyspectrum}
\nu \ =\ 2 (\nu _{+}+n)\quad \text{or}\quad 
\nu \ =\ 2 (\nu _{-}+n)\quad \quad
\text{with }\ n\in \mathbb{N}\ \ .
\end{equation}
Let us remark that in the special case $\nu _{+}=1-\nu _{-}$, this
solution becomes the exact spectrum for real $\nu $. 
The spectrum in eq.\ (\ref{btlargeenergyspectrum}) is just the union
of the spectra of two Liouville theories with 
boundary parameters $\nu _{+}$ and $\nu _{-}$. Indeed we obtain an
average level spacing of $\delta \nu =1$. In our analysis we found it 
hard to control the case when $\nu $ comes very close to an integer 
number, but the comparison with the semi-classical expectations 
tells us that we already found all eigenvalues.  
\smallskip

Now we want to analyse the limit $\Delta \to -\infty $. If we set 
$\nu =i\omega $, we have to study the limit $\omega \to \infty$.
From the condition (\ref{btspectrum}) we obtain 
\begin{equation}\label{negspeccond}
\frac{1+\xi ^{2}_{i\omega } (\lambda   )}{1-\xi ^{2}_{i\omega }
(\lambda   )}\ = \ 
i \ \frac{1-\tan \pi \nu_{+}\tan \pi \nu _{-}}{\tan \pi 
\nu _{+}+\tan \pi \nu _{-}} + \mathcal{O} (e^{-\pi \omega })\ \ .
\end{equation}
For large values of $\omega $ we can approximate $\xi_{i\omega }
(\lambda  )$ 
using eq.\ (\ref{approxxi}) and Stirling's formula,
\[
\xi_{i\omega } (\lambda  ) \ =\ -i\, e^{i\omega (\log \frac{\lambda }{4\omega
^{2}}+2)} \ \big(1+\mathcal{O} (1/\omega ) \big) \ \ .
\]
If we substitute this expression into eq.\ (\ref{negspeccond}), we 
find that the spectrum of large negative $\Delta =-\omega ^{2}$ is
determined by the equation
\[
\omega (\log (4 \omega^{2}/\lambda ) -2)+\mathcal{O} (1/\omega )\ =\ \pi n+
\text{arctan}\frac{\tan \pi \nu _{+}\tan \pi \nu _{-}-1}{\tan \pi \nu
_{+}+\tan \pi \nu _{-}} 
\]
with $n\in \mathbb{N}$. The spacing $\delta \omega $ between two
solutions of this equation takes the form  
\[
\delta \omega \ \approx \ \pi  \big( \log (4\omega ^{2}/\lambda ) \big)^{-1}
\ \ .
\]
When we translate this into the energy difference $\delta\Delta=2
\sqrt{-\Delta }\, \delta \omega  $ between two eigenvalues, we 
recover the semi-classical result~(\ref{semclassspectrum}). 

\section{Conclusions and open problems} 

Above we have analysed the minisuperspace model of rolling and 
bouncing tachyons, i.e.\ of time-like Liouville and sine-Gordon 
theory. In contrast to the more complicated field theories, their 
minisuperspace toy models are easily treated directly, without 
performing a Wick rotation from the corresponding Euclidean 
background. The results we obtained are certainly very suggestive 
of several features in the field theory models. In particular, we 
expect that time-like (boundary) Liouville theory comes with one 
real parameter $\nu_0$ which describes some boundary condition in 
the far future $x_0 = \infty$. The latter must be imposed because 
in an exponentially unbounded potential, particles and strings 
reach infinity in finite world-sheet time. Technically, the 
parameter comes in through the need to make the generator of 
world-sheet time translations self-adjoint. We have made similar 
statements about the time-like (boundary) sine-Gordon model only 
that in this case a 2-parameter family is predicted from the 
minisuperspace analysis. Each member of this family can be argued 
to possess a purely discrete spectrum of conformal weights. Finally, 
we saw that our two toy models contain a discrete set of states 
with positive eigenvalue $\Delta$. It remains to be seen whether 
the associated field theories also possess primaries of arbitrarily 
large conformal weight, or whether their spectrum gets truncated.
\smallskip 

The construction of the full field theory models is certainly  
a very interesting open problem. Progress in this direction could 
possibly be made along different lines. Inspired by our observation 
that  minisuperspace quantities in our rolling tachyon models are 
obtained from the Wick-rotated theory by a replacement $\exp(-2 \pi 
\omega) \rightarrow -\zeta^{\nu_0}(\omega)$, one might try to start 
from the known expression for bulk 3-point functions in the 
Wick-rotated $c=1$ Liouville theory \cite{Schomerus:2003vv} (see 
also \cite{Strominger:2003fn} for earlier expressions of the bulk 
2-point functions) and proceed to their $\nu_0$-dependent `components' 
by some kind of `unitarisation'. 
Another, maybe more promising approach could pass through path 
integral representations of the time-like model. Using arguments 
as in \cite{Goulian:1991qr}, this might lead to modified screening integrals 
and then ultimately to new $\nu_0$-dependent solutions of time-like 
Liouville theory, very much along the lines of \cite{Teschner:1995yf,
Fateev:2000ik,Zamolodchikov:2001ah}. We plan to come back to these 
problems in the near future. 
\medskip 

While our results above allow to draw rather obvious and well 
motivated conclusions for the involved  conformal field theories, 
it seems less clear that the parameters $\nu_0$ or $\nu_\pm$ will 
also show up in string theory amplitudes. In string theory, the 
Hamiltonian of the underlying world-sheet theory is only used 
to formulate the physical state condition, i.e.\ it appears in 
the form of a constraint. Even though the usual techniques to solve 
such constraints do also exploit self-adjointness, this may not 
stand up as a firm argument against other scenarios in which 
string theory amplitudes involve some averaging over different 
CFT backgrounds. Actually, there are several examples of 
dynamical processes in string theory which are believed to 
end in a `mixed' final state. This is particularly well 
established for the condensation of an open string tachyon 
which can cause a single brane to decay into a final  
configuration containing several lower dimensional branes. 
Examples of mixed final states that arise from the condensation 
of closed string tachyons can be found e.g.\ in \cite{Harvey:2001wm}.  
Clues on the role of the parameters $\nu_0$ and $\nu_\pm$ in 
string theory might come from more thorough investigations of 
quantum field theories with time dependent masses or from the 
duality between closed strings in a flat background and open 
strings on decaying branes (see e.g.\ \cite{Karczmarek:2003xm} 
for some recent study of this duality). We believe that this 
issue deserves further investigation.    
\bigskip \bigskip 

\noindent
{\bf Acknowledgements:} We would like to thank R.\ Kashaev for
a discussion that initiated this work and for sharing results 
of a collaboration with L.D.\ Faddeev and M.\ Volkov. Moreover, 
we are indebted to A.Yu.\ Alekseev, M.\ Berkooz, K.\ Fredenhagen,
H.\ Grosse, S.\ Ribault, A.\ Sen, J.\ Teschner and T.\ Thiemann 
for very useful conversations.      

\begin{appendix}
\section{Completeness}
Our aim in this appendix is to show that the set of eigenfunctions
$\psi _{\nu _{n}}^{\nu _{0}},\psi _{\omega  }^{\nu _{0}}$ (see eqs.\
(\ref{eig<},\ref{eig>})) is complete, i.e.\ that eq.\
(\ref{completeness}) is fulfilled. 

We start with the integral
\[
I (\alpha ,\alpha' )\ =\ \int_{0}^{\infty } d\omega \ \overline{\psi _{\omega
}^{\nu _{0}}} (\alpha ) \psi _{\omega }^{\nu _{0}} (\alpha ')  
\]
where $\alpha =\sqrt{\lambda }e^{x_{0}}$, $\alpha '=\sqrt{\lambda }e^{x_{0}'}$.
We insert the expression (\ref{eig>}) for $\psi _{\omega }^{\nu _{0}}$
and obtain
\begin{align*}
I (\alpha ,\alpha ')\ =\ 2\int_{0}^{\infty }d\omega \ \frac{\omega
}{\sinh 2\pi \omega }\big( & J_{2i\omega } (\alpha )J_{-2i\omega }
(\alpha ') + J_{-2i\omega } (\alpha )J_{2i\omega } (\alpha' )\\
& + \zeta^{\nu _{0}} (-\omega )J_{-2i\omega } (\alpha )J_{-2i\omega } (\alpha
')+\zeta^{\nu _{0}} (\omega )J_{2i\omega } (\alpha )J_{-2i\omega } (\alpha ')\big)\ \ .
\end{align*}
Here, $\zeta ^{\nu _{0}} (\omega )$ is defined in eq.\
(\ref{defofzeta}). 
We expect this integral to be a distribution, so we introduce an extra
factor $1/ (\varepsilon^{2} \omega ^{2}+1)$ in the integral to regularize
it. We allow $\varepsilon $ to be complex with positive real part. At
the end we shall take the limit $\varepsilon\to 0$.  

Let us assume that $\alpha \geq \alpha '$. We evaluate the integral by
rewriting it first as an integral over the whole real line, 
\[
I_{\varepsilon } (\alpha ,\alpha ')\ =\ 2\int_{-\infty }^{\infty
}d\omega \ \frac{\omega }{\sinh 2\pi \omega }\big( J_{2i\omega }
(\alpha )J_{-2i\omega } (\alpha ')+\zeta^{\nu _{0}} (-\omega )J_{-2i\omega }
(\alpha )J_{-2i\omega } (\alpha ')\big)\frac{1}{\varepsilon^{2}\omega
^{2}+1 } \ \ .
\]
By a careful analysis of the asymptotic behaviour of $J_{\nu }$
for large order $|\nu |$ we find that we can close the contour in the
upper half-plane without any contribution from the semi-circle at
infinity. Using Cauchy's theorem, we can evaluate our integral by the
sum of residues of poles in the upper half-plane. Note that there are 
no poles at the zeroes of $\sinh 2\pi \omega $. The only poles come
from $\zeta^{\nu _{0}} (-\omega )$ at $\omega =i (\nu _{0}+n), n\in
\mathbb{N}^{*}$ and from our regulator at $\omega =i/\varepsilon $. We
choose $\varepsilon $ s.t.\ the poles do not meet.  
The contribution from the residues at $\omega =i (\nu _{0}+n)$ is
\[
2\pi i\sum (\text{Residues})\ =\  - 4\sum_{n=0}^{\infty } (\nu _{0}+n)J_{2 (\nu _{0}+n)}
(\alpha )J_{2 (\nu _{0}+n)} (\alpha ')  \frac{1}{1-\varepsilon^{2}
(n+\nu _{0})^{2} }\ \ .
\]
In the limit $\varepsilon \to 0$ this approaches the contribution from
the discrete spectrum.

It remains to evaluate the residue at $\omega =i/\varepsilon $ in
the limit $\varepsilon\to 0 $. This is a straightforward exercise
in playing with the asymptotic behaviour of Bessel functions at large
order (see e.g.\ \cite{Watson:book}). We obtain at the end
\[
I (\alpha ,\alpha ') +\sum_{n=0}^{\infty }\overline{\psi _{\nu
_{n}}^{\nu _{0}}} (\alpha )\psi _{\nu _{n}}^{\nu _{0}} (\alpha ')\ =\
\lim_{\varepsilon\to 0 } \frac{1}{\varepsilon } e^{-2|\log
(\alpha /\alpha ')|/\varepsilon } \ \ . 
\]
The right hand side is an expression for the delta-distribution
$\delta (\log (\alpha /\alpha '))=\delta (x_{0}-x_{0}')$. This is what
we wanted to show.
\end{appendix}

\end{document}

%% file: rt.pstex_t
\begin{picture}(0,0)%
\special{psfile=rt.pstex}%
\end{picture}%
\setlength{\unitlength}{3947sp}%
\begingroup\makeatletter\ifx\SetFigFont\undefined
\def\x#1#2#3#4#5#6#7\relax{\def\x{#1#2#3#4#5#6}}%
\expandafter\x\fmtname xxxxxx\relax \def\y{splain}%
\ifx\x\y   
\gdef\SetFigFont#1#2#3{%
  \ifnum #1<17\tiny\else \ifnum #1<20\small\else
  \ifnum #1<24\normalsize\else \ifnum #1<29\large\else
  \ifnum #1<34\Large\else \ifnum #1<41\LARGE\else
     \huge\fi\fi\fi\fi\fi\fi
  \csname #3\endcsname}%
\else
\gdef\SetFigFont#1#2#3{\begingroup
  \count@#1\relax \ifnum 25<\count@\count@25\fi
  \def\x{\endgroup\@setsize\SetFigFont{#2pt}}%
  \expandafter\x
    \csname \romannumeral\the\count@ pt\expandafter\endcsname
    \csname @\romannumeral\the\count@ pt\endcsname
  \csname #3\endcsname}%
\fi
\fi\endgroup
\begin{picture}(4059,4348)(1864,-4664)
\put(5386,-1531){\makebox(0,0)[lb]{\smash{\SetFigFont{12}{14.4}{rm}\fbox{$\nu_0=\frac{3}{4}$}}}}
\put(4688,-496){\makebox(0,0)[lb]{\smash{\SetFigFont{12}{14.4}{rm}$\lambda e^{2x_0}$}}}
\put(5872,-3567){\makebox(0,0)[lb]{\smash{\SetFigFont{12}{14.4}{rm}$x_0$}}}
\put(3661,-4508){\makebox(0,0)[lb]{\smash{\SetFigFont{12}{14.4}{rm}-5}}}
\put(3953,-2506){\makebox(0,0)[lb]{\smash{\SetFigFont{12}{14.4}{rm}5}}}
\put(3923,-1509){\makebox(0,0)[lb]{\smash{\SetFigFont{12}{14.4}{rm}10}}}
\put(3433,-3609){\makebox(0,0)[lb]{\smash{\SetFigFont{12}{14.4}{rm}-1}}}
\put(3083,-3609){\makebox(0,0)[lb]{\smash{\SetFigFont{12}{14.4}{rm}-2}}}
\put(2733,-3609){\makebox(0,0)[lb]{\smash{\SetFigFont{12}{14.4}{rm}-3}}}
\put(2383,-3609){\makebox(0,0)[lb]{\smash{\SetFigFont{12}{14.4}{rm}-4}}}
\put(2033,-3609){\makebox(0,0)[lb]{\smash{\SetFigFont{12}{14.4}{rm}-5}}}
\put(5229,-3609){\makebox(0,0)[lb]{\smash{\SetFigFont{12}{14.4}{rm}4}}}
\put(4879,-3609){\makebox(0,0)[lb]{\smash{\SetFigFont{12}{14.4}{rm}3}}}
\put(4529,-3609){\makebox(0,0)[lb]{\smash{\SetFigFont{12}{14.4}{rm}2}}}
\put(5579,-3609){\makebox(0,0)[lb]{\smash{\SetFigFont{12}{14.4}{rm}5}}}
\put(4179,-3609){\makebox(0,0)[lb]{\smash{\SetFigFont{12}{14.4}{rm}1}}}
\put(3968,-564){\makebox(0,0)[lb]{\smash{\SetFigFont{12}{14.4}{rm}$\Delta$}}}
\end{picture}

%% file: psv.pstex_t
\begin{picture}(0,0)%
\special{psfile=psv.pstex}%
\end{picture}%
\setlength{\unitlength}{3947sp}%
\begingroup\makeatletter\ifx\SetFigFont\undefined
\def\x#1#2#3#4#5#6#7\relax{\def\x{#1#2#3#4#5#6}}%
\expandafter\x\fmtname xxxxxx\relax \def\y{splain}%
\ifx\x\y   
\gdef\SetFigFont#1#2#3{%
  \ifnum #1<17\tiny\else \ifnum #1<20\small\else
  \ifnum #1<24\normalsize\else \ifnum #1<29\large\else
  \ifnum #1<34\Large\else \ifnum #1<41\LARGE\else
     \huge\fi\fi\fi\fi\fi\fi
  \csname #3\endcsname}%
\else
\gdef\SetFigFont#1#2#3{\begingroup
  \count@#1\relax \ifnum 25<\count@\count@25\fi
  \def\x{\endgroup\@setsize\SetFigFont{#2pt}}%
  \expandafter\x
    \csname \romannumeral\the\count@ pt\expandafter\endcsname
    \csname @\romannumeral\the\count@ pt\endcsname
  \csname #3\endcsname}%
\fi
\fi\endgroup
\begin{picture}(2555,6183)(2626,-7106)
\put(2710,-1073){\makebox(0,0)[lb]{\smash{\SetFigFont{12}{14.4}{rm}$\Delta$}}}
\put(3639,-7017){\makebox(0,0)[lb]{\smash{\SetFigFont{12}{14.4}{rm}$\Gamma(0,\Delta)/h$}}}
\put(4097,-6761){\makebox(0,0)[lb]{\smash{\SetFigFont{12}{14.4}{rm}0}}}
\put(2943,-6761){\makebox(0,0)[lb]{\smash{\SetFigFont{12}{14.4}{rm}-4}}}
\put(2710,-2088){\makebox(0,0)[lb]{\smash{\SetFigFont{12}{14.4}{rm}10}}}
\put(2792,-4067){\makebox(0,0)[lb]{\smash{\SetFigFont{12}{14.4}{rm}0}}}
\put(2806,-3070){\makebox(0,0)[lb]{\smash{\SetFigFont{12}{14.4}{rm}5}}}
\put(2724,-5065){\makebox(0,0)[lb]{\smash{\SetFigFont{12}{14.4}{rm}-5}}}
\put(3744,-6761){\makebox(0,0)[lb]{\smash{\SetFigFont{12}{14.4}{rm}-1}}}
\put(4631,-6761){\makebox(0,0)[lb]{\smash{\SetFigFont{12}{14.4}{rm}2}}}
\put(4364,-6761){\makebox(0,0)[lb]{\smash{\SetFigFont{12}{14.4}{rm}1}}}
\put(4898,-6761){\makebox(0,0)[lb]{\smash{\SetFigFont{12}{14.4}{rm}3}}}
\put(3210,-6761){\makebox(0,0)[lb]{\smash{\SetFigFont{12}{14.4}{rm}-3}}}
\put(3477,-6761){\makebox(0,0)[lb]{\smash{\SetFigFont{12}{14.4}{rm}-2}}}
\put(2626,-6068){\makebox(0,0)[lb]{\smash{\SetFigFont{12}{14.4}{rm}-10}}}
\end{picture}

%% file: bt.pstex_t
\begin{picture}(0,0)%
\special{psfile=bt.pstex}%
\end{picture}%
\setlength{\unitlength}{3947sp}%
\begingroup\makeatletter\ifx\SetFigFont\undefined
\def\x#1#2#3#4#5#6#7\relax{\def\x{#1#2#3#4#5#6}}%
\expandafter\x\fmtname xxxxxx\relax \def\y{splain}%
\ifx\x\y   
\gdef\SetFigFont#1#2#3{%
  \ifnum #1<17\tiny\else \ifnum #1<20\small\else
  \ifnum #1<24\normalsize\else \ifnum #1<29\large\else
  \ifnum #1<34\Large\else \ifnum #1<41\LARGE\else
     \huge\fi\fi\fi\fi\fi\fi
  \csname #3\endcsname}%
\else
\gdef\SetFigFont#1#2#3{\begingroup
  \count@#1\relax \ifnum 25<\count@\count@25\fi
  \def\x{\endgroup\@setsize\SetFigFont{#2pt}}%
  \expandafter\x
    \csname \romannumeral\the\count@ pt\expandafter\endcsname
    \csname @\romannumeral\the\count@ pt\endcsname
  \csname #3\endcsname}%
\fi
\fi\endgroup
\begin{picture}(4185,6280)(998,-6198)
\put(3848,-151){\makebox(0,0)[lb]{\smash{\SetFigFont{12}{14.4}{rm}$2\lambda\cosh 2 x_0$}}}
\put(998,-1186){\makebox(0,0)[lb]{\smash{\SetFigFont{12}{14.4}{rm}\fbox{$\nu_-=\frac{1}{4}$}}}}
\put(4426,-1186){\makebox(0,0)[lb]{\smash{\SetFigFont{12}{14.4}{rm}\fbox{$\nu_+=\frac{3}{4}$}}}}
\put(2754,-5146){\makebox(0,0)[lb]{\smash{\SetFigFont{12}{14.4}{rm}-10}}}
\put(2851,-4134){\makebox(0,0)[lb]{\smash{\SetFigFont{12}{14.4}{rm}-5}}}
\put(3128,-1155){\makebox(0,0)[lb]{\smash{\SetFigFont{12}{14.4}{rm}10}}}
\put(3128,-2161){\makebox(0,0)[lb]{\smash{\SetFigFont{12}{14.4}{rm}5}}}
\put(1211,-3248){\makebox(0,0)[lb]{\smash{\SetFigFont{12}{14.4}{rm}-5}}}
\put(1559,-3248){\makebox(0,0)[lb]{\smash{\SetFigFont{12}{14.4}{rm}-4}}}
\put(1907,-3248){\makebox(0,0)[lb]{\smash{\SetFigFont{12}{14.4}{rm}-3}}}
\put(2255,-3248){\makebox(0,0)[lb]{\smash{\SetFigFont{12}{14.4}{rm}-2}}}
\put(2603,-3248){\makebox(0,0)[lb]{\smash{\SetFigFont{12}{14.4}{rm}-1}}}
\put(4744,-3248){\makebox(0,0)[lb]{\smash{\SetFigFont{12}{14.4}{rm}5}}}
\put(3701,-3248){\makebox(0,0)[lb]{\smash{\SetFigFont{12}{14.4}{rm}2}}}
\put(4049,-3248){\makebox(0,0)[lb]{\smash{\SetFigFont{12}{14.4}{rm}3}}}
\put(4397,-3248){\makebox(0,0)[lb]{\smash{\SetFigFont{12}{14.4}{rm}4}}}
\put(3353,-3248){\makebox(0,0)[lb]{\smash{\SetFigFont{12}{14.4}{rm}1}}}
\put(3143,-68){\makebox(0,0)[lb]{\smash{\SetFigFont{12}{14.4}{rm}$\Delta$}}}
\put(5116,-3234){\makebox(0,0)[lb]{\smash{\SetFigFont{12}{14.4}{rm}$x_0$}}}
\end{picture}